\RequirePackage{fix-cm}
\documentclass[conference]{IEEEtran}
\IEEEoverridecommandlockouts

\usepackage{cite}
\usepackage{amsmath,amssymb,amsfonts}
\usepackage{algorithmic}
\usepackage{graphicx}
\usepackage{textcomp}
\usepackage{xcolor}
\usepackage[flushmargin]{footmisc} 

\usepackage{algorithmic}
\usepackage{textcomp}
\usepackage{subfigure}
\usepackage{xspace}
\usepackage{mathtools}
\usepackage{enumitem}
\usepackage{multirow}
\usepackage{adjustbox}
\usepackage{tikz}
\usepackage{siunitx}
\usepackage{verbatim}
\usepackage{mdframed}
\usepackage{listings}
\usepackage{xurl}
\usepackage{booktabs}

\def\BibTeX{{\rm B\kern-.05em{\sc i\kern-.025em b}\kern-.08em
    T\kern-.1667em\lower.7ex\hbox{E}\kern-.125emX}}

\begin{document}

\title{Exploring the Jupyter Ecosystem: An Empirical Study of Bugs and Vulnerabilities\thanks{This work is supported by the Natural Sciences and Engineering Research Council of Canada (NSERC-DG), grant No. 2024-00347.}
}
\author{
\IEEEauthorblockN{Wenyuan Jiang}
\IEEEauthorblockA{
\textit{ETH Z\"urich}\\
Z\"urich, Switzerland \\
wenyjiang@student.ethz.ch}
\and
\IEEEauthorblockN{Diany Pressato}
\IEEEauthorblockA{
\textit{Concordia University}\\
Montreal, Canada \\
diany.pressato@mail.concordia.ca	}
\and
\IEEEauthorblockN{Harsh	Darji}
\IEEEauthorblockA{
\textit{University of Alberta}\\
Camrose, Canada \\
hdarji@ualberta.ca}
\and
\IEEEauthorblockN{Thibaud	Lutellier}
\IEEEauthorblockA{
\textit{University of Alberta}\\
Camrose, Canada \\
lutellie@ualberta.ca}
}

\IEEEpubid{
\makebox[\columnwidth]{979-8-3315-9147-2/25/\$31.00~\copyright~2025 IEEE\hfill}
\hspace{\columnsep}
\makebox[\columnwidth]{}
}

\definecolor{emerald}{rgb}{0.31, 0.78, 0.47}

\newcommand{\rawnumber}[1]{{\color{black}#1}}


\newcommand{\SE}{SE\xspace} 
\newcommand{\nonSEfull}{Educational\xspace} 
\newcommand{\nonSEshort}{Edu.\xspace} 


\newcommand{\interrateraggreementNb}{88\%\xspace} 

\newcommand{\interrateraggreementPy}{85\%\xspace} 

\newcommand{\numSEactiveNb}{376\xspace} 

\newcommand{\numnonSEactiveNb}{525\xspace} 

\newcommand{\numSEactivePy}{608\xspace} 

\newcommand{\numnonSEactivePy}{110\xspace} 


\newcommand{\totalGitHubSE}{{8,647}\xspace}

\newcommand{\totalGitHubNonSE}{{28,302}\xspace}

\newcommand{\totalGitHubNotebooks}{{36,949}\xspace}

\newcommand{\totalforkSE}{{10,929}\xspace}

\newcommand{\totalactiveforkSE}{{1,090}\xspace}

\newcommand{\percentactiveforkSE}{{10\%}\xspace}

\newcommand{\totalforkNONSE}{{15,300}\xspace}

\newcommand{\totalactiveforkNONSE}{{964}\xspace}

\newcommand{\percentactiveforkNONSE}{{6\%}\xspace}



\newcommand{\totalGitHubPython}{{\rev{NOT CORRECT 117,175}}\xspace} 
\newcommand{\totalGitHubPythonChanges}{884,650\xspace}  
\newcommand{\totalGitHubNotebookChanges}{587,381\xspace}

%
\newcommand{\medianStargazersGitHubNotebooks}{{1,238}\xspace} 
\newcommand{\medianStargazersGitHubPython}{{1,868}\xspace} 
\newcommand{\medianForksGitHubNotebooks}{{250}\xspace} 
\newcommand{\medianForksGitHubPython}{{394}\xspace} 
\newcommand{\medianIssuesGitHubNotebooks}{{15}\xspace} 
\newcommand{\medianIssuesGitHubPython}{{28}\xspace} 


%
\newcommand{\meanCyclomaticComplexityGitHubNotebooks}{{27.9}\xspace} 
\newcommand{\meanCyclomaticComplexityGitHubPython}{{23.3}\xspace} 
\newcommand{\medianCyclomaticComplexityGitHubNotebooks}{{18}\xspace} 
\newcommand{\medianCyclomaticComplexityGitHubPython}{{7}\xspace} 


%
\newcommand{\meanCommitCountNbSE}{{889}\xspace} 
\newcommand{\sdCommitCountNbSE}{{3,001}\xspace} 

\newcommand{\meanCommitCountPySE}{{1,391}\xspace} 
\newcommand{\sdCommitCountPySE}{{5,983}\xspace}  

\newcommand{\meanCommitCountNbnonSE}{{483}\xspace} 
\newcommand{\sdCommitCountNbnonSE}{{1,222}\xspace} 

\newcommand{\meanCommitCountPynonSE}{{362}\xspace} 
\newcommand{\sdCommitCountPynonSE}{{746}\xspace}  
\newcommand{\medianCommitIntervalNbSE}{{6.6}\xspace} 
\newcommand{\medianCommitIntervalPySE}{{7.3}\xspace} 

\newcommand{\medianCommitIntervalNbnonSE}{{6.8}\xspace} 
\newcommand{\medianCommitIntervalPynonSE}{{7.4}\xspace} 
\newcommand{\avgCommitIntervalNbPySE}{{14}\xspace} 

\newcommand{\meanCommitAuthorsNbSE}{{29}\xspace} 
\newcommand{\meanCommitAuthorsNbnonSE}{{22}\xspace} 

\newcommand{\meanCommitAuthorsPySE}{{34}\xspace} 
\newcommand{\meanCommitAuthorsPynonSE}{{18}\xspace} 

\newcommand{\medianCommitMsgLenNbSE}{{90}\xspace} 
\newcommand{\medianCommitMsgLenPySE}{{97}\xspace} 
\newcommand{\medianCommitMsgEntropyNbSE}{{4.09}\xspace} 
\newcommand{\medianCommitMsgEntropyPySE}{{4.23}\xspace} 

\newcommand{\medianCommitMsgLenNbnonSE}{{49}\xspace} 
\newcommand{\medianCommitMsgLenPynonSE}{{64}\xspace} 
\newcommand{\medianCommitMsgEntropyNbnonSE}{{3.90}\xspace}  
\newcommand{\medianCommitMsgEntropyPynonSE}{{4.10}\xspace} 


\newcommand{\geoLocationCount}{{958}\xspace}  
\newcommand{\geoCountryCount}{{78}\xspace}  
\newcommand{\geoContributionCountAll}{{9,806}\xspace} 
\newcommand{\geoContributionCountNorthAmerica}{{3,781}\xspace} 
\newcommand{\geoContributionCountAsia}{{2,846}\xspace} 
\newcommand{\geoContributionCountEurope}{{2,608}\xspace} 
\newcommand{\geoContributionCountSouthAmerica}{{157}\xspace} 
\newcommand{\geoContributionCountAfrica}{{90}\xspace} 
\newcommand{\maxAccountAgeYearNotebook}{{14}\xspace} 


\newcommand{\totalKaggleNotebooks}{{1,170}\xspace} 
\newcommand{\totalKaggleNotebooksParsed}{{1,038}\xspace} 


\newcommand{\notebookRepoFileCountPython}{{55,302}\xspace}
\newcommand{\notebookRepoFileCountDocument}{{51,198}\xspace}
\newcommand{\notebookRepoFileCountDatafile}{{16,849}\xspace}
\newcommand{\notebookRepoFileConfig}{{13,094}\xspace}
\newcommand{\notebookRepoFileCXX}{{10,000}\xspace}
\newcommand{\notebookRepoFileJS}{{6,214}\xspace}
\newcommand{\notebookRepoFileMatlab}{{1,951}\xspace}
\newcommand{\notebookRepoFileJava}{{1,778}\xspace}

\newcommand{\NPMICoherence}{{0.07}\xspace}
\newcommand{\CVCoherenceAllActiveRepos}{{0.64}\xspace}
\newcommand{\NPMICoherenceAllActiveRepos}{{0.06}\xspace}
\newcommand{\numAllActiveHeaders}{{261,471}\xspace}
\newcommand{\numAllAciveOutliers}{{104,468}\xspace}

\newcommand{\CVCoherenceSE}{{0.69}\xspace}
\newcommand{\NPMICoherenceSE}{{0.10}\xspace}
\newcommand{\numSETopics}{{1,114}\xspace}
\newcommand{\numSEHeaders}{{48,949}\xspace}
\newcommand{\numSEOutliers}{{11,393}\xspace}

\newcommand{\CVCoherenceNonSE}{{0.66}\xspace}
\newcommand{\NPMICoherenceNonSE}{{0.06}\xspace}
\newcommand{\numNonSETopics}{{734}\xspace}
\newcommand{\numNonSEHeaders}{{212,522}\xspace}
\newcommand{\numNonSEOutliers}{{83,724}\xspace}

\newcommand{\numManualTopics}{{18}\xspace}

\newcounter{finding}
\newcommand{\finding}[1]{\refstepcounter{finding}
	\begin{mdframed}[linecolor=gray,roundcorner=12pt,backgroundcolor=gray!15,linewidth=3pt,innerleftmargin=2pt, leftmargin=0cm,rightmargin=0cm,topline=false,bottomline=false,rightline=false]
	 #1
      \end{mdframed}
}

\maketitle
\IEEEpubidadjcol

\begin{abstract}
Background. Jupyter notebooks are one of the main tools used by data scientists. Notebooks include features (configuration scripts, markdown, images, etc.) that make them challenging to analyze compared to traditional software. 
As a result, existing software engineering models, tools, and studies do not capture the uniqueness of Notebook's behavior. 

Aims. This paper aims to provide a large-scale empirical study of bugs and vulnerabilities in the Notebook ecosystem.

Method. We collected and analyzed a large dataset of Notebooks from two major platforms. Our methodology involved quantitative analyses of notebook characteristics (such as complexity metrics, contributor activity, and documentation) to identify factors correlated with bugs. Additionally, we conducted a qualitative study using grounded theory to categorize notebook bugs, resulting in a comprehensive bug taxonomy. Finally, we analyzed security-related commits and vulnerability reports to assess risks associated with Notebook deployment frameworks.

Results. Our findings highlight that configuration issues are among the most common bugs in notebook documents, followed by incorrect API usage. Finally, we explore common vulnerabilities associated with popular deployment frameworks to better understand risks associated with Notebook development.

Conclusions. This work highlights that notebooks are less well-supported than traditional software, resulting in more complex code, misconfiguration, and poor maintenance.
\end{abstract}

\section{Introduction}
Data science plays an increasingly pivotal role in driving global economic advancements.
Positioned at the convergence of diverse scientific domains, data scientists, while adept in their respective scientific disciplines, often lack extensive familiarity with software development and reliability practices.
This knowledge gap may inadvertently lead to the introduction of software bugs and vulnerabilities within data science applications.
Such implications underscore growing concerns regarding the reliability, security, and overall dependability of data science software, emphasizing the need for a deeper understanding of software engineering principles within the data science community. 

For example, 2022 saw the first ransomware attack on Jupyter Notebook~\cite{morag2022threats}, one of the most popular web applications for data science programming. 
The attack encrypted the users' notebooks and demanded a ransom payment for decryption. 
The success of this attack was attributed to misconfigured software security settings, emphasizing the critical importance of addressing and rectifying notebook bugs to safeguard against security breaches.

Jupyter Notebook has become an industry standard for writing code related to data exploration, analysis, and machine learning~\cite{jetbrain2022report}. 
These documents intertwine source code (e.g., Python, Matlab, or C) with descriptive markdown text that can include equations or media content. 
Furthermore, these documents showcase the dynamic outputs of the executed code, including variable values, tables, and graphs, providing a comprehensive and interactive platform for coding, documentation, and data visualization.

\begin{figure}[ht]
\centering
\includegraphics[width=0.42\textwidth]{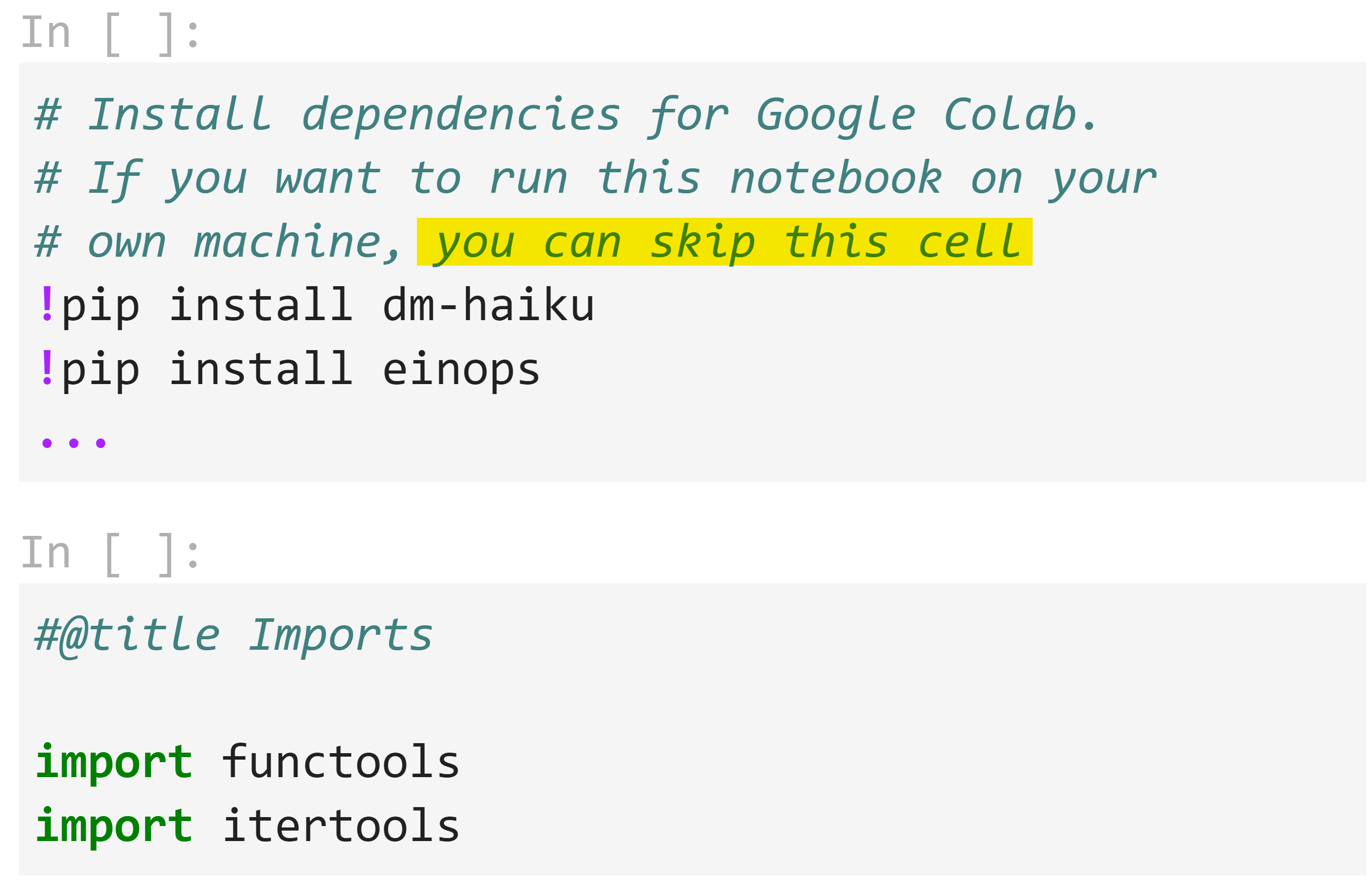}
\caption{A Notebook where the first cell may be skipped.}
\label{fig:cell_exec_number}
\end{figure}

Unlike traditional programming environments where code must be executed sequentially, Jupyter Notebooks allow users to run cells individually, enabling a more interactive and exploratory coding experience. 
Data scientists take full advantage of this feature as shown in Figure~\ref{fig:cell_exec_number}. 
In this real-world notebook from the Google DeepMind repository~\cite{deepmindIntro}, developers create specific configuration cells that should only be run if working on Colab. If users work on their local machines, they should skip the execution of this cell.

The non-sequential execution capability of Notebooks, while proposing an interactive coding environment, presents a challenge for software engineering tools. 
These unique features, combined with the accessibility of notebooks to non-experts, increase the probability of bugs within notebooks.
In addition, even when proficient notebook users employ debugging tools, these solutions may fall short in addressing the distinctive characteristics inherent to notebooks. 
As a consequence, their efficacy in assisting users in enhancing notebook reliability may be compromised. 

Despite recent efforts focused on reproducibility~\cite{pimentel2021understanding,wang2020assessing,wang2020better,zhu2021restoring} and the establishment of best practices~\cite{quaranta2022eliciting,nahar2022collaboration,settewong2022visualize}, the notebook ecosystem is not yet well understood, and an in-depth study of notebooks and their associated bugs is needed. In particular, given recent security incidents involving notebooks~\cite{morag2022threats}, understanding the prevalence and nature of security vulnerabilities has become especially critical.

We address this gap through quantitative and qualitative analyses of notebook characteristics and changes.  First, we gather a dataset of Notebook files and commits extracted from active GitHub repositories. We then quantitatively analyze the correlation between notebook characteristics and the frequency of bugs  (Section~\ref{sec:explore}), develop a taxonomy of notebook bugs (Section~\ref{sec:taxonomy}), and analyze security-related commits and vulnerabilities to identify prevalent security issues (Section~\ref{sec:security}).

This paper makes the following contributions:

\smallskip\noindent\textbf{Contribution 1:} An empirical study of Jupyter Notebook characteristics and their correlation to bugs. We provide insights into factors most strongly associated with notebook bugs, guiding future tool development and best practices.

\smallskip\noindent\textbf{Contribution 2:} A comprehensive taxonomy of Jupyter Notebook bugs derived from empirical analysis. This taxonomy helps researchers and practitioners better understand prevalent issues and their root causes.

\smallskip\noindent\textbf{Contribution 3:} An analysis of security vulnerabilities within Jupyter Notebook deployment frameworks. We highlight critical security risks, emphasizing the need for robust security practices in notebook infrastructure.

\section{Motivation \& Background }

\subsection{Motivation}
Despite the growing adoption of Jupyter Notebooks across data science, machine learning, and scientific computing, the software engineering community still lacks a comprehensive understanding of the quality challenges they introduce. Unlike traditional software, notebooks support nonlinear execution, mixing of code and documentation, dynamic outputs, and minimal testing infrastructure. These unique characteristics can lead to hard-to-detect bugs and poorly maintained projects. Moreover, the ease of use and accessibility of notebooks attracts a wide range of users, many of whom may lack formal software development training, increasing the likelihood of defects and misconfigurations. These challenges call for a systematic investigation of bugs within the notebook ecosystem.

To address this gap, we formulate three research questions. RQ1 explores which notebook characteristics correlate with higher bug frequency, aiming to identify risk factors and inform quality-assurance practices. RQ2 seeks to create a taxonomy of notebook-specific bugs through qualitative analysis, helping researchers and tool builders understand the types of errors that commonly arise in this environment. Finally, RQ3 investigates the security vulnerabilities present in notebook deployment frameworks, an increasingly urgent concern given recent high-profile attacks and the widespread use of notebooks in production pipelines. Together, these questions provide a view of notebook reliability, from structural contributors to systemic risks.

\subsection{Background}

A \textbf{Jupyter Notebook} is a document that enables users to write and execute code interactively, one cell at a time. In addition to code, users can include markdown text, equations, media, and hyperlinks—organizing content into either code cells or markdown cells. A typical Jupyter Notebook consists of three main components:

Jupyter Notebooks support multi-language code execution, allowing users to run \textbf{code cells} in any order. A code cell includes executable code and  configuration scripts.

\textbf{Markdown cells} allow users to add text or media, enhancing code cells with explanations and visual context.

Jupyter notebooks are executable documents that include output. 
Each code cell has its \textbf{output cell} that may contain text, traceback contents, graphs, images, videos, audio clips, and any output generated from a code cell.
\section{Methodology}
\label{sec:approach}

\begin{figure*}[!t]
\centering
\includegraphics[width=0.80\textwidth]{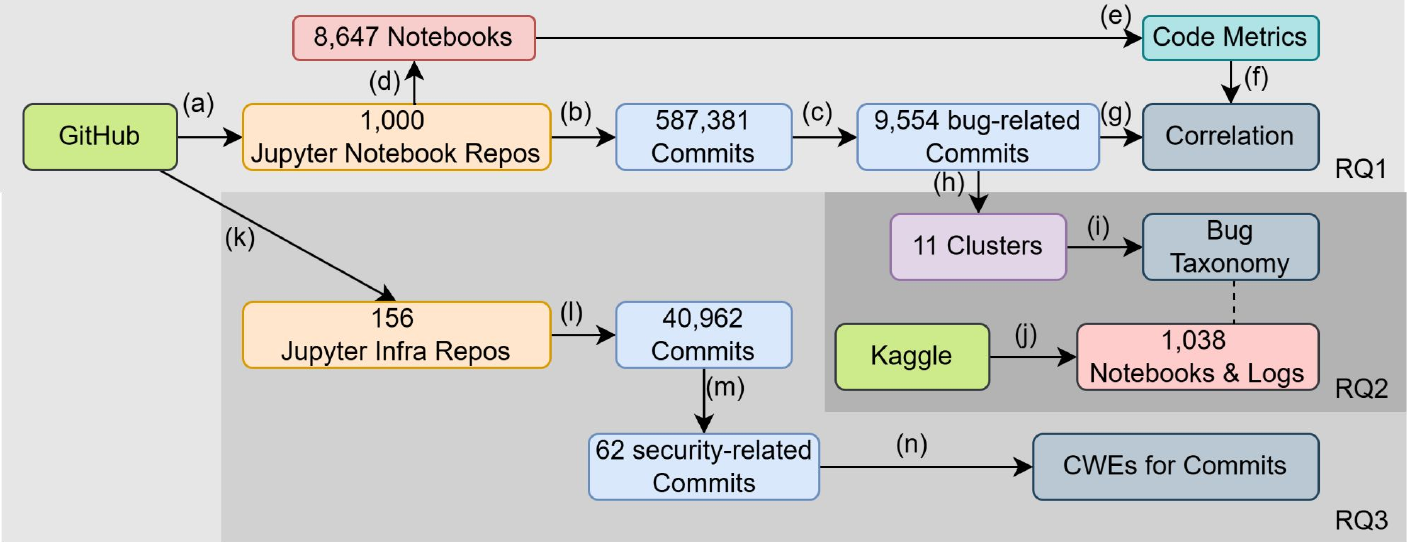}

\caption{Overview of the data collection and processing pipelines for each research question. }
\label{fig:pipeline}
\end{figure*}

To guide our investigation, we formulate the following three research questions:
\textbf{RQ1:} Which characteristics of Jupyter Notebooks correlate with a higher frequency of bugs?,
\textbf{RQ2:} What are the most common types of bugs in Jupyter Notebook documents?
\textbf{RQ3:} What are the most common security vulnerabilities in Notebook deployment frameworks?

Our approach consists of four main steps, as illustrated in Fig.~\ref{fig:pipeline}. 
First, we mine active Jupyter Notebook repositories to extract notebook characteristics and code changes (Section~\ref{sec:data}). We then conduct an empirical study to identify code and project features that correlate with the presence of bugs, addressing RQ1 (Section~\ref{sec:study}). To answer RQ2, we perform a qualitative analysis to develop a taxonomy of common bugs in Jupyter Notebooks (Section~\ref{sec:approchtax}). Finally, we investigate security issues in notebook deployment frameworks such as JupyterLab and JupyterHub, addressing RQ3 (Section~\ref{sec:security}). Altogether, these steps provide a comprehensive view of bugs and vulnerabilities across the Jupyter Notebook ecosystem.

\subsection{Data Extraction}
\label{sec:data}
For our empirical study, we chose two orthogonal sources of data. 
Our first source comes from open-source GitHub repositories, while our second source comes from the Kaggle platform. We selected GitHub because it hosts a vast and diverse array of notebook-based projects spanning numerous application domains, reflecting real-world development practices. Kaggle was chosen due to its competitive environment, which encourages high-quality documentation and robust coding practices, thus complementing our GitHub dataset by providing a different perspective on notebook usage.

\smallskip\noindent\textbf{GitHub Project Selection:}
Using the GitHub API, we download the top 1,000 public repositories labeled with the ``Jupyter Notebook'' language tag, sorted by star ratings (Step (a) in Figure~\ref{fig:pipeline}).
Then, to increase the quality of our dataset, we removed inactive projects. 
We consider a project active if it has received a GitHub event (other than a watch event) within the last year (as of February 2024). 
Furthermore, since anyone can post personal projects on GitHub, and Jupyter Notebooks are used for a wide variety of less relevant projects, including books or blogs, two of the co-authors manually checked the README and the project descriptions of the repositories to discard non-development-related projects. 
This selection of relevant projects resulted in an inter-rater agreement of \interrateraggreementNb and disagreements were settled by a third co-author. After this filtering process, we obtain \numSEactiveNb active Notebook repositories.

\smallskip\noindent\textbf{GitHub Notebook Characteristics and Change Extraction:} 
To answer RQ1, we extracted the characteristics of Notebook files and projects (Steps (d) and (e)), including complexity metrics, natural language metrics, 
and contributor metrics. 

Extracting changes in Jupyter notebooks is challenging due to the complexity of the notebook format. 
Unlike traditional code files, Jupyter notebooks store content in a structured manner that includes not only code but also rich text, images, and interactive elements. 
Moreover, Jupyter notebooks store metadata and cell outputs within the same file, complicating the differencing process. 

Consequently, conventional differential algorithms (e.g., git diff) fail to isolate and represent changes in Jupyter notebooks. 
These algorithms tend to capture alterations in both structure and output, leading to substantial discrepancies. 
For instance, in cases where images are generated as output, the git diff algorithm may register thousands of lines as changed, reflecting the regeneration of a new image, even if only a few lines of code were modified.

To retrieve the code changes for each file in one commit, we extract the file before and after the corresponding commit. 
We then concatenate all Python source code cells in each notebook and parse them into a Python abstract syntax tree (AST).
The ASTs are then serialized and compared.
The results highlight changes in the code AST, which is more useful than standard line-based diff algorithms. Overall, we extracted~\totalGitHubNotebookChanges unique changes in Jupyter Notebook files.

\smallskip\noindent\textbf{Kaggle Notebook Selection:}
The Kaggle notebooks' selection was based on Kaggle's ranking system. We employed a systematic algorithm to extract notebooks and their latest log files from the first 20 pages (which roughly mapped to the 1,000 most popular files) of Kaggle's notebook repository. Overall, we gathered~\totalKaggleNotebooksParsed Kaggle's competition notebooks and execution logs (Step (j)).
These notebooks, submitted for competitive challenges, span a diverse range of applications, providing further insights into varied Jupyter notebook usage practices. 
Our focus on Kaggle competition notebooks was driven by the perceived higher quality in terms of code documentation and log file maintenance. 
This conclusion was drawn from an analysis of the substantial incentives offered in Kaggle competitions, including significant cash rewards and opportunities for developers to showcase their skills in machine learning and data science. 

We further parse the logs, looking for the keywords `Traceback' and `Error' to identify bugs and map them back to code cells in the original Kaggle notebook. Finally, we extract the execution time of each file from the logs.

\subsection{RQ1 Settings}
\label{sec:study}

\smallskip\noindent\textbf{Motivation:} 
Understanding which characteristics of Jupyter Notebooks correlate with a higher frequency of bugs in Notebook files is essential for improving the reliability, reproducibility, and maintainability of data science workflows. 
Jupyter Notebooks are widely used in exploratory data analysis, machine learning model development, and research, but their interactive and flexible nature introduces challenges that increase the likelihood of bugs. 
These bugs may arise from factors such as code complexity, lack of documentation, lower developer experience, dependence on external libraries and data, or uniqueness of the domain of applications.

Identifying the specific characteristics that contribute to these issues allows for the development of better practices, tools, and guidelines to mitigate bug introduction, improving both individual productivity and collaboration in data science teams. 
By exploring these correlations, we can uncover patterns that are not only specific to Notebooks but may also apply to other interactive computing environments.
We describe below how we identify bug-related commits and how we measure correlations between bugs and notebook features.

\smallskip\noindent\textbf{Identifying Bug-Related Commits:} 
\label{sec:bugfiltering}
From the dataset of changes extracted from GitHub Notebook repositories (Section~\ref{sec:data}), we identify bug-relate commits following previous work~\cite{wang2016automatically,lutellier2020coconut} by considering a commit as bug-related if its related message contains any of the keywords ``fix'', ``bug'', ``patch'' but does not contain the keywords ``rename,'' ``merge,'' ``clean-up,'' or ``refactor.'' Overall, we extracted 9,554 bug-related commits (step (c)).

\smallskip\noindent\textbf{Correlation With General Project Characteristics:} 
From our dataset of \totalGitHubSE notebooks and 587,381 unique changes, we extracted metadata and characteristics that may be correlated to bugs. Specifically, we investigate different sets of characteristics covering code complexity, natural language, and contributor metadata. 

To assess code complexity, we examined five metrics including the number of functions, lines of code, libraries imported and code blocks in the Notebook, and  the cyclomatic complexity of the file. These features were selected because they are recognized as indicators of code complexity, which is often linked to the likelihood of bugs in traditional software.

We also measure three natural language metrics, including code-to-markdown line ratio, length of commit messages, and the entropy of these messages. Code-to-markdown ratio is important as markdown serves as the primary medium for developers to incorporate natural language into Notebooks, impacting readability and maintenance. The length and entropy of commit messages, while not direct measures of clarity, provide indirect insights into how contributors document changes.

Finally, we look at the number of contributors working on the same file, the number of commits on a given file, and the age of their GitHub accounts at the time of contribution. The number of contributors working on the same file reflects the level of collaboration, which can introduce coordination challenges and potential inconsistencies. The age of the GitHub accounts of the contributors at the time of the contribution is a proxy for their experience~\cite{bao2019large,eluri2021predicting}. These metrics were chosen to capture collaboration and developer experience, factors frequently highlighted as influencing software quality.
We measure the correlation between these metrics and the number of bugs in a given file, we use the Pearson Correlation.

\subsection{RQ2 Settings:}
\label{sec:approchtax}

\smallskip\noindent\textbf{Motivation:} RQ2 aims at creating a taxonomy of bugs in Notebooks. While general software defect taxonomies exist, they often fail to account for the unique features of notebooks. By systematically categorizing common bugs in this environment, we aim to uncover patterns that are unique to notebook-based workflows. This is essential for building debugging tools, informing best practices, and supporting educational initiatives tailored to the notebook ecosystem.
 
 We follow the grounded theory methodology to do a qualitative analysis of the bugs in Jupyter Notebooks. This process led us to the building of a Jupyter Notebook bug taxonomy. 
We describe below our main process.

\smallskip\noindent\textbf{Data Collection Stage:} Our main data comes from the two sources as described in Section~\ref{sec:data}. 
For this specific study, we only focus on the~\totalGitHubNotebookChanges GitHub notebook changes and on logs from Kaggle notebooks that contain any traceback.

\smallskip\noindent\textbf{Open Coding Stage:}
The open coding stage is a foundational step in our qualitative analysis, aimed at identifying distinct types of bug fixes in Jupyter Notebooks. To manage the large volume of commits and focus our manual analysis, we applied an automated clustering strategy to segment the dataset into meaningful groups of similar bug-fixing commits. This structured approach helped ensure diversity in the types of bugs we reviewed while maintaining feasibility.

We began by applying a mixed heuristic sampling process based on commit messages and AST-level code changes. Because Jupyter Notebooks are stored in JSON format, traditional line-based diff tools produce noisy results that include metadata and output differences irrelevant to actual code changes. 
For example, Listing~\ref{lst:codeChange} displays a change where a fix requires a single line change (swap between \texttt{input} and \texttt{input-1} parameters, lines 10 and 11). In the Notebook format, the diff is more complex to parse since the developer also changed the input file and this change resulted in the change of a very large tensor (in the data field) since such data is stored in the notebook. Additionally, metadata fields are also updated with the \texttt{execution\_count} of every cell in the Notebook being updated. This makes correctly isolating bug-fixing changes very challenging in the Notebook environment. To address this, we extracted the Python source code cells info from the diff and used AST parsing to represent changes.

We then parsed these source code cells into abstract syntax trees (ASTs) and applied a structural diffing algorithm to identify semantic changes. This AST-based comparison reveals more meaningful transformations than line-based diffs—for example, distinguishing the insertion of control structures like an if condition, which might appear trivial in a line diff but indicates deeper logical modifications.

\begin{lstlisting}[caption={Example of a 1-line fix resulting in a change complex to parse.}\label{lst:codeChange}]
"source": [
1 - "(Path(`train/<@\textcolor{purple}{002844.jpg}@>'), [`train'])"
2 + "(Path(`train/<@\textcolor{teal}{008663.jpg}@>'),\n", 
3 +     [`car', 'person'])
4 [...]
5 "data": { "text/plain": [
6 -   "tensor([<@\textcolor{purple}{-1.0028, [...], -3.6006}@>],\n"
7 +   "tensor(<@\textcolor{teal}{[ 2.0258, [...],  1.6073}@>],\n"
8 [...]
9 "source": [
10 - " return  where<@\textcolor{purple}{(1-inputs, inputs}@>).mean()"
11 + " return  where<@\textcolor{teal}{(inputs, 1-inputs}@>).mean()"
12 [...]
 \end{lstlisting}

To characterize each bug fix, we select eight metrics describing properties such as the size and type of changes, as well as boolean flags indicating whether the modification occurs within a loop, condition, constant, and so on.
Principal Component Analysis (PCA) reduces these metrics to two principal components, which are then clustered with DBSCAN. 
This procedure initially yields 8 clusters that achieve a silhouette score~\cite{rousseeuw1987silhouettes} of 0.91, indicating high-quality clustering (where negative values suggest misclassified instances, 0 implies overlapping clusters, and 1 is the ideal value). 
To cover all relevant scenarios, 3 additional clusters are introduced to account for non-code edits (e.g., markdown cell updates), unparsable file changes, and commits that lack bug-fixing keywords. 
In total, we produced 11 clusters for further qualitative analysis.

\smallskip\noindent\textbf{Axial Coding:} Three co-authors then each randomly sample 30 or 50 bugs from different categories (based on availability), manually analyze the changes, commit messages and associated issues or logs and come up with categories and descriptions for each bug. 

\smallskip\noindent\textbf{Selective Coding and Saturation:} 
After this initial sampling process, the three co-authors had an open discussion to consolidate their categorization and bug descriptions. 
Once the discussion is done, 30 or 50 additional bugs per person are sampled again. 
After three iterations and \rawnumber{230} bugs manually investigated, none of the authors found additional insights or categories, ending the search.

\subsection{RQ3 Settings:}
\label{sec:security}

\noindent\textbf{Motivation:} Jupyter Notebook infrastructures such as JupyterHub, Jupyter Server, and JupyterLab have grown in popularity as core components of data science and interactive computing workflows. However, their widespread adoption also increases the threat surface for potential attacks. We observed when collecting data that the yearly count of Common Vulnerabilities and Exposures (CVEs) reported for software within the Jupyter Notebook ecosystem from 2015 to 2024 significantly increased. For example, the number of security reports in the NVD referencing Jupyter notebooks doubled between 2023 and 2024. This trend emphasizes the necessity of examining how security vulnerabilities manifest in real-world notebook deployment frameworks.

\smallskip\noindent\textbf{Security-Related Commit Selection:}
To study security issues in notebook deployment frameworks, we investigate all open-source repositories (156) from the three most popular GitHub organizations for notebook infrastructure: jupyter-hub, jupyter-server, and jupyterlab. We cloned the repositories and extracted all the commits in the main branch, resulting in a dataset of 40,962 commits. To obtain commits related to security issues, we use a two-stage filtering pipeline described below. 

First, following previous work~\cite{zhou2017automated}, we use regular-expression-based (regex) method (available in our replication package) on commit messages to filter out commits that are not relevant to security issues.
To avoid including commits generated by automation tools, we also excluded commits that have a commit message longer than 1,000 characters. After this first stage, we obtained 400 security-related commits. 

Regex filtering often leads to a relatively high false positive rate due to its limitation in understanding the semantics of the commit message. 
Inspired by the widespread use of large language models (LLMs) in software engineering, our second stage includes an LLM-based method for refining the results produced by the Regex filtering (full prompt in our replication package).
We leveraged a state-of-the-art LLM, DeepSeek-V3~\cite{deepseekai2024deepseekv3technicalreport}, for filtering due to its performance on software engineering tasks. The LLM is used to match a given commit message to an entry of the ``2024 CWE Top 25 Most Dangerous Software Weaknesses''\cite{mitreTop25}.
We used one-shot prompting on the LLM. The prompt included the description of the task, the output format, an example commit message and its paired expected output (a CWE ID), as well as the commit message to evaluate. Out of the previous 400 commits, 323 were matched with a CWE entry.

Hallucination is a problem of LLMs. To mitigate this issue, we manually checked the 323 commits for errors in the CWE mapping. We found that commits from one of the repositories (`zero-to-jupyterhub-k8s') only contained irrelevant commits from automatic vulnerability scans. After eliminating the automated commit scans, we ended up with 66 security-related commits. We further manually annotated to validate the LLM's output according to the CWE website's guidelines, reducing our total number of security-related commits to 62. After this annotation, we grouped the CWEs according to their parent pillar CWE in the CWE-1000 view to consolidate our results. 
\section{RQ1 Results}
\label{sec:explore}

To address RQ1—identifying characteristics of Jupyter Notebooks that correlate with a higher frequency of bugs—we analyzed the correlation between various notebook attributes and the number of bug-related commits.

\begin{figure}[ht]
\centering
\includegraphics[width=\columnwidth]{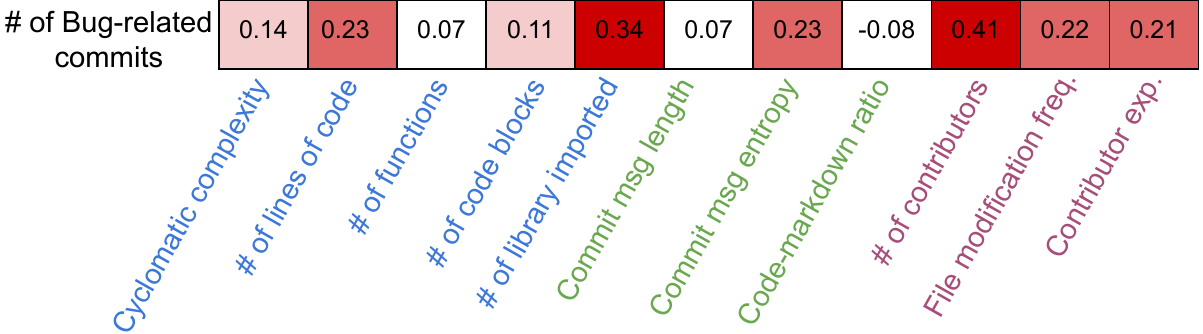}
\caption{Pearson Correlation coefficients between the number of bug-related commits and Notebook metrics. White, light red, medium red, and dark red indicate negligible, weak, medium, and strong correlations. Blue, green, and purple coding represent code-related, natural language, and metadata metrics.}
\label{fig:correlation_matrix}
\end{figure}

Figure~\ref{fig:correlation_matrix} presents the Pearson correlation values that illustrate the relationships between the different notebook characteristics and bug-related commits. Blue characteristics (e.g., Cyclomatic complexity) are code-related metrics, green ones are natural language metrics while the purple ones are metadata metrics related to file contributors. All the metrics describe file-level characteristics. For example, the ``Commit msg length'' measured the average commit message length of all commits modifying a specific notebook file. Pearson coefficients close to 1 (or -1) indicate a strong correlation between two features, while a coefficient close to 0 indicates a lack of correlation between features. For readability, we also colored the Pearson coefficients, with stronger red colors representing stronger correlations. For example, the first column of the Figure indicates that there is a weak positive (0.14) correlation between bug-related commits modifying a file and its cyclomatic complexity. Several key findings emerge from this analysis. We divide our analysis between code, natural language, and contributors metrics.

\smallskip\noindent\textbf{Code Metrics:} Our results on code metrics are diverse. Files importing many libraries and large files are more likely to be modified by bug-related commits (Pearson coefficients of 0.34 and 0.23), reinforcing the idea that longer, complex files are more bug-prone.
We found no correlation between function count and bug-related commits, indicating that the use (or non-use) of functions in Notebooks doesn't significantly impact the number of bugs in Jupyter Notebooks. The number of code blocks and cyclomatic complexity of the file play a slightly greater role (Pearson coefficient of 0.11 and 0.14) but still represent only a weak correlation.

\smallskip\noindent\textbf{Natural Language Metrics:}
Two of the natural language metrics (commit message length and code-to-markdown ratio) are not significantly correlated with bug-related commits. This is surprising as commit metadata has previously been shown to perform well for bug prediction~\cite{zhou2022simple} despite its simplicity. Code-to-markdown ratio not being significant tends to indicate that markdown plays a different role than code comments (i.e., it is not used to describe the code) as code comments in traditional software have been shown to be related to software quality and bug proneness~\cite{ibrahim2012relationship}. This suggests that markdown cells, despite being another way to mix natural language and source code, do not replace the need for comments. 

\smallskip\noindent\textbf{Contributor Metrics:}
Contributor metrics are the metrics most correlated with bug-related commits. Specifically, the number of contributors metric is strongly correlated, indicating that more contributors tend to modify buggy files, possibly due to increased complexity in coordination and merging changes. This high correlation compared to other metrics may also indicate that Notebooks are not as convenient for teamwork as more traditional source code. Similarly, frequently modified files are more prone to bugs, possibly due to instability from frequent changes. Finally, experienced users are more likely to work on files containing more bug-related commits, possibly because they are trusted with fixing issues.

\smallskip\finding{\textbf{RQ1 Summary:} The main correlation on bugs is related to contributors' behavior and characteristics more than any code complexity or natural language metrics. This suggests that software engineering tools made for sharing Notebooks and working as a team (e.g., version control systems) are not well-suited for Jupyter Notebooks.}

\section{RQ2 Results: Bug Taxonomy}
\label{sec:taxonomy}

In this section, we follow grounded theory to generate a taxonomy of bugs' root causes in Jupyter Notebooks in order to understand the most common types of bugs present in the Notebook ecosystem. The resulting root cause categories of our study are described below.

\smallskip\noindent\textbf{Incorrect configuration:}
Notebooks share the same enormous and ever-growing ecosystem as Python does, and therefore share the tedious configuration process for many libraries. 
There are two major configuration problems that cause bugs.
The first one is path-related. 
When users need to specify the location of resources, a path or url needs to be given.
Without a structured configuration management tool, the common practice for a notebook is to manually fill in some path/url string, which may lead to problems when there is a change to the running environment.
Code Snippet~\ref{lst:bug_misConfig1} is an example where a misconfigured path is updated.

\begin{lstlisting}[caption={Example of an Incorrect configuration bug}\label{lst:bug_misConfig1}]
- path_data = '<@\textcolor{purple}{../../../data/}@>'
+ path_data = '<@\textcolor{teal}{../../../assets/data/}@>'
\end{lstlisting}

The other type of configuration bugs comes from versions of both the Python runtime and libraries.
Unlike the package manager, npm, in the Javascript ecosystem, the default Python package manager, pip, does not force to specify a version when installing a library.
This leads to incompatibility problems when some of the libraries a notebook depends on have undergone major updates.
For example, one bug in a Colab notebook is caused by not specifying the Tensorflow version~\cite{dl-colab-notebooks}. 
The fix depends on the `magic command' feature provided by both the Jupyter Notebook ecosystem and the Colab environment, as is shown in Code Snippet~\ref{lst:bug_misConfig2}.
\begin{lstlisting}[caption={Example of an Incorrect configuration bug}\label{lst:bug_misConfig2}]
+ <@\textcolor{teal}{\%tensorflow\_version 1.x}@>
\end{lstlisting}

Magic functions, also referred to as `IPython magics', are specialized commands within Jupyter notebooks designed to offer convenient shortcuts and additional functionalities. Prefixed with a \% symbol, these commands enable users to execute various operations beyond standard Python syntax. Throughout our research, we noticed frequent misuse of magic functions. This was often rooted in typographical errors, minor syntax mistakes, or the selection of an inappropriate magic function for a particular task. While magic functions provide a valuable means to enhance Jupyter notebook functionality, instances of misapplication can result in unintended errors, potentially compromising the overall integrity of the notebook.

\smallskip\noindent\textbf{Data shape/structure:}
One of the major use cases for Jupyter Notebooks is the data science field, where users need to manipulate different shapes of data, ranging from unstructured text, semi-structured JSON files to different shapes of matrices and tensors.
As the core of notebooks, python provides various syntactic sugar for convenient manipulation of one or multiple data items in a very concise statement, which may take traditional languages like Java several or tens of statements to do. 
Third-party libraries like numpy and pandas made this coding style even more popular, which also complicates the semantics of these syntactic sugars.
Bugs caused by incorrect handling of data structures are often seen. 
This is made worse by the lack of static type checking in Python interpreters.
For example, a bug in a tutorial was caused by the misuse of certain shapes of tensors~\cite{pythoncode-tutorials}. 

The fix of this bug, as is shown in Code Snippet~\ref{lst:bug_dataShape}, features the flexibility of Python syntax for tensors.

\begin{lstlisting}[caption={Example of a Data shape/structure bug}\label{lst:bug_dataShape}]
- mae = data.inverse_transform(<@\textcolor{purple}{mae.reshape(1,-1)}@>)[0][0]
+ mae = data.inverse_transform(<@\textcolor{teal}{[[mae]]}@>)[0][0]
\end{lstlisting}

\smallskip\noindent\textbf{API misuse:}
Notebooks span very diverse topics, with many domains relying on specific APIs such as TensorFlow, Pandas, website's REST APIs, etc. 
Many bugs in Jupyter Notebooks are related to incorrect usage of such API, resulting in incorrect function calls, missing or incorrect input parameters, or misunderstanding in return types.

\begin{lstlisting}[caption={Example of an Incorrect API bug}\label{lst:bug_incorrectAPI}]
- return <@\textcolor{purple}{vsm\_leaves\_phi}@>(text, yelp_lookup, np_func)
+ return <@\textcolor{teal}{vsm\_phi}@>(text, yelp_lookup, np_func)
\end{lstlisting}

Additional instances of API misuse arise from customized methods created by Jupyter Notebook users, stemming from the absence of automated refactoring tools available to Notebook developers. 
Code Snippet~\ref{lst:bug_incorrectAPI} depicts an example of such a bug~\cite{cs224u}.
The developers altered the method's name from \lstinline{vsm_leaves_phi} to 
\lstinline{vsm_phi}. However, due to the lack of automatic refactoring tools, the author did not update all calls to this function. 
In Jupyter Notebooks, the presence of the original function name, \lstinline{vsm_leaves_phi}, persists in memory until the user reruns the cell defining the function and no crash is observed. The issue will only manifest itself if a user reruns the modified cell.

\smallskip\noindent\textbf{Incorrect syntax:} In this category, we observe struggles linked to Python's syntax for object-oriented programming, such as incorrect usage of the keyword self. 
This type of bug happened in the popular fast.ai framework\cite{fastai2Commit2}. This may be connected to the fact that Jupyter Notebook developers rarely use any classes or functions, and may not be comfortable with oriented object programming features.

\begin{lstlisting}[caption={Example of an Incorrect syntax bug}\label{lst:bug_incorrectSyntax}]
- self.opt.set_hyper(<@\textcolor{purple}{opt.hypers}@>[0]['lr']*self.mult_lr)
+ self.opt.set_hyper(<@\textcolor{teal}{self.opt.hypers}@>[0]['lr']*self.mult_lr)
\end{lstlisting}

\smallskip\noindent\textbf{Wrong logic:}
Code Snippet~\ref{lst:bug_wrongLogic} shows an example of Wrong Logic in the HuggingFace notebook repository\cite{WrongLogic}. 
In the original implementation, the author missed cases where the answer span partially overlapped with the context. 

\begin{lstlisting}[caption={Example of a Wrong logic bug}\label{lst:bug_wrongLogic}]
- if offset[context_start][0] > <@\textcolor{purple}{end\_char}@> \
-    or offset[context_end][1] < <@\textcolor{purple}{start\_char}@>:
+ if offset[context_start][0] > <@\textcolor{teal}{start\_char}@> \
+    or offset[context_end][1] < <@\textcolor{teal}{end\_char}@>:
\end{lstlisting}

Another type of ``Wrong Logic'' bug occurs when developers pick an incorrect algorithm. 
For example, we found a bug in the Google DeepMind repository~\cite{incorrectWeights} where the weights for computing the optical flow were incorrectly set up.

\smallskip\noindent\textbf{Non-determinism:}
Probabilistic features (e.g., the random standard library) are widely used in modern programs to implement simulation and machine learning algorithms.
One common practice for ensuring reproducibility is to use pseudo-random functions and explicitly specify the random seed.
However, due to the complexity of the Notebook ecosystem, even when users specify the random seed, there are cases when unexpected random and non-deterministic bugs happen.

Code Snippet~\ref{lst:bug_rng} shows an example of a randomness bug from fastai2~\cite{randomnessBug}.
When initializing the random number generator, the author of the notebook didn't set the seed according to an already-seeded random source, therefore causing the bug that even when users explicitly set a random seed, the results are not reproducible. 
To fix the bug, the contributors passed a random number generated from the user-defined seed as the seed for the new random number generator, making the new random number generator deterministic. 

\begin{lstlisting}[caption={Example of a Random-related bug}\label{lst:bug_rng}]
- self.rng = random.Random()
+ self.rng = random.Random(<@\textcolor{teal}{random.randint(0,2**32-1)}@>)
\end{lstlisting}

\smallskip\noindent\textbf{Exception/Error/Log/Debugging:}
The cell-based execution model of Jupyter Notebooks makes it difficult for users to reason about the current execution context when an exception happens. 
Also, the cell-based output mechanism adds complexity to the classical standard output mechanism assumed by Python, leading to traditional logging and debugging utilities that are difficult to use correctly.

\smallskip\noindent\textbf{Errors in Test Code and Assertions:}
The flexibility of cell execution and the convenience of result visualization in Jupyter Notebooks make test frameworks neither necessary nor compatible.
One practice of testing a piece of code in notebooks is by creating another cell or directly adding assertion statements to the target code. 
This simple testing method could be used incorrectly by users and cause bugs. 

\smallskip\noindent\textbf{Resource management:} 
These bugs encompass issues related to the allocation and utilization of computing resources, such as the number of GPUs allocated, input sizes, and batch sizes. Improper handling of these aspects can lead to suboptimal execution, performance bottlenecks, or even system failures. For instance, we found a bug in the fastai2~\cite{ResourceManagement} repository where developers allocated an excessively large input size, overwhelming available memory, causing the notebook to crash. Identifying and addressing these issues enhances the reproducibility of analyses and contributes to the overall robustness of Jupyter Notebooks.

\smallskip\noindent\textbf{Incomplete code:} There are several instances of incomplete code in Jupyter Notebooks. Unlike traditional programming languages, incomplete code in notebooks often arises from educational or tutorial contexts. However, we consider incomplete code to be a bug because it leads to runtime errors or prevents users from properly executing or understanding subsequent notebook cells. Thus, incomplete code negatively impacts notebook functionality and usability, warranting its inclusion in our bug taxonomy and qualitative analysis.

\smallskip\noindent\textbf{Undeclared variables \& Typos:}
Many bugs we found would be considered minor, yet ended up in Jupyter Notebooks of popular organizations such as fast.ai, or in the "Data Science on AWS" textbook. These bugs highlight the lack of static analysis and linting in the default Notebook web interface.  

\smallskip\noindent\textbf{Documentation mistakes:} Mistakes in markdown may be considered bugs since they will affect the understanding of the complete document. For example, a markdown cell may describe a specific mathematical formula that is then implemented in Python. If that formula is incorrectly described, this is a bug as the document becomes incorrect. This is analogous
to mistakes in comments being bugs~\cite{zhai2020cpc,tan2015code,tan2012tcomment,tan2007icomment}.

\smallskip\noindent\textbf{Kaggle Notebooks and Logs Analysis:}
In inspecting \totalKaggleNotebooksParsed Kaggle notebooks, \rawnumber{89} instances (\rawnumber{9\%}) exhibiting traceback contents were identified, signifying errors within the logs. These errors were categorized into four primary types: Compiler Error (\rawnumber{59} notebooks), Logic Error (\rawnumber{5} notebooks), Configuration Error (\rawnumber{20} notebooks), and API Misuse (\rawnumber{5} notebooks).

\textbf{Incorrect configuration errors} underscore the importance of configuration management, including dependencies, environment settings, file paths, and configuration errors. \textbf{Kaggle Compiler errors} are often attributed to syntax issues or runtime anomalies within the codebase. \textbf{API errors} arise from issues with interfacing external services or libraries. Finally, \textbf{Logic errors} are caused by flawed implementations.

These findings from Kaggle align closely with our GitHub-based taxonomy of bugs. In both datasets, configuration issues and API misuse were among the most frequently observed root causes. While the Kaggle dataset contained additional compiler errors, both platforms demonstrated the presence of logic errors, reinforcing the generality of these bug categories. This overlap validates the robustness of our taxonomy across diverse notebook usage contexts and highlights consistent pain points for notebook users regardless of the platform.

\smallskip\finding{\textbf{RQ2 Summary:}
The most common bug root causes in Jupyter Notebooks fall into 12 main categories, with incorrect configuration, data shape mistakes, API misuse, incomplete code, wrong logic, and documentation errors being the most common root causes. Kaggle-based observations confirm our observations on GitHub with configuration issues, API misuse, and wrong logic being the main root causes of bugs.}
\section{RQ3 Results: Notebook Security}
\label{sec:security-results}
RQ3 investigates security issues in Jupyter Notebooks. While we didn't find any security issues in Notebook documents, deployment frameworks contain security vulnerabilities that might make running Notebook unsafe. For example, CVE-2024-43805 refers to a security vulnerability in the Jupyter-lab deployment environment where opening a maliciously crafted notebook allows an attacker to perform a cross-site scripting attack. Below, we present our findings on security-related commits in Jupyter Notebook infrastructures, drawing on the methodology described in Section~\ref{sec:security}.

Table~\ref{table:cwe-count} summarizes the number of commits under each CWE pillar in the CWE-1000 view, with each CWE pillar item indicated by its CWE ID and name. For example, the first line of the table indicates that we found 19 commits in Jupyter Notebook deployment frameworks that fix vulnerabilities related to CWE-693, Protection Mechanism Failure.

The frequency of CWE types in the selected commits follows a long-tail distribution.
The most frequently observed CWE was Protection Mechanism Failure (CWE-693), followed by Improper Access Control (CWE-284) and Improper Control of a Resource Through Its Lifetime (CWE-664). 
These CWEs often arise in web-based systems and align with the Jupyter Notebook architecture, which relies on HTTP services and browsers. 
Less frequent CWEs include Command Injection (CWE-77) and Use of Known Vulnerable Component (CWE-1395), both of which are also typical in web services.

\begin{table}[t]
\caption{Number of commits related to each CWE pillar entry.}
	\centering
	\begin{tabular}{r l}
    \toprule
	\textbf{\# of} & \textbf{Common Weakness Enumeration ID} \\ 
    \textbf{Commits} & \textbf{(CWE ID)} \\ 
    \midrule
    19&CWE-693: Protection Mechanism Failure\\
    17&CWE-284: Improper Access Control\\
    12&CWE-664: Improper Control of a Resource  \\
    &Through its Lifetime\\
    8&CWE-710: Improper Adherence to Coding\\
    &Standards\\
    4&CWE-707: Improper Neutralization\\
    1&CWE-691: Insufficient Control Flow Management\\
    1&CWE-703: Improper Check or Handling of \\
    &Exceptional Conditions\\
    \bottomrule
	\end{tabular}
	
	\label{table:cwe-count}
\end{table}

\begin{lstlisting}[caption={Example of CSRF bug in the jupyterhub project}\label{lst:csrf}]
jupyterhub/handlers/base.py
+ <@\textcolor{teal}{clear\_xsrf\_cookie\_kwargs = \{}@>
+    <@\textcolor{teal}{key: value for key, value in }@>
+    <@\textcolor{teal}{            self.settings.get('xsrf\_cookie\_kwargs', \{\})}@>
+    <@\textcolor{teal}{if key in \{"path", "domain"\}\}}@>
+    <@\textcolor{teal}{\}}@>
self.clear_cookie('_xsrf',
-    <@\textcolor{purple}{**self.settings.get('xsrf\_cookie\_kwargs', \{\}),}@>
+    <@\textcolor{teal}{**clear\_xsrf\_cookie\_kwargs,}@> )
\end{lstlisting}

Within Protection Mechanism Failure, the most recurrent vulnerability concerns Cross-Site Request Forgery, accounting for 14 of the 19 relevant commits. Listing~\ref{lst:csrf} shows an example of such a bug in the JupyterHub framework.

\begin{lstlisting}[caption={An Access Control bug in the nbgitpuller project}\label{lst:acl}]
nbgitpuller/handlers.py
class LegacyInteractRedirectHandler(IPythonHandler):
+   @<@\textcolor{teal}{web.authenticated}@>
    def get(self):
\end{lstlisting}

Access control is also a significant bug in notebook infrastructure.
Improper Access Control (CWE-284) was often addressed by bolstering authentication for notebook infrastructures that were previously unsecured. 
For example, a vulnerability in the nbgitpuller was fixed by ``making sure that all endpoints are authenticated'', which improves the missing access control for notebook infrastructure (See Listing~\ref{lst:acl}).

Other web-based vulnerabilities included Cross-Site Scripting (XSS) due to insufficient input sanitization (CWE-707).
Listing~\ref{lst:xss} shows an example of fixing an XSS vulnerability. Here, the addition of the ``autoescape'' option will automatically escape special characters, preventing the vulnerability.

\begin{lstlisting}[caption={Example of an XSS bug in the notebook project}\label{lst:xss}]
notebookapp.py
- <@\textcolor{purple}{jenv\_opt = jinja\_env\_options}@>
+ <@\textcolor{teal}{jenv\_opt = \{"autoescape": True\}}@>
+ <@\textcolor{teal}{jenv\_opt.update(jinja\_env\_options}@>
\end{lstlisting}

These findings are consistent with recent CVE reports on the Notebook infrastructures. Among the 60 CVE reports we collected from 2015 to 2024, the majority of HIGH or CRITICAL rated vulnerabilities relate to web or access control issues, highlighting similar security concerns observed in our commit analysis. 
CVE reports also give insights beyond the infrastructure repos in our study, as many CVEs are rooted in the ecosystem of Jupyter Notebooks like servers and plugins.

\begin{lstlisting}[caption={Security Relax for Single User Server}\label{lst:secRelax}]
jupyterhub/singleuser.py:
+ <@\textcolor{teal}{@classmethod}@>
+ <@\textcolor{teal}{def validate\_security(cls, app, ssl\_options=None):}@>
+     <@\textcolor{teal}{return}@>
\end{lstlisting}

\smallskip\noindent\textbf{Trade-off between usability and security:} 
Most security-related commits involved fixes for discovered vulnerabilities. 
However, a subset of commits relaxed security settings to simplify configuration in common single-user scenarios. 
For instance, Listing~\ref{lst:secRelax} displays a security change~\cite{securityRelax} that overwrites the validate\_security method to suppress TLS-related security warnings for single-user servers. 
These changes are not advisable for publicly accessible deployments; however, they address usability challenges often encountered by individual or non-professional users.
In the examples above, both TLS and cross-origin configuration are known to be difficult and messy in a non-standard web production scenario.
These commits make configuration easier for out-of-the-box use of notebooks with a trade-off between usability and security.

\smallskip\noindent\textbf{Security implications:}
Although Jupyter Notebook infrastructures were originally designed for single-user, local usage, our analysis shows that they now face a range of common web-related vulnerabilities (e.g., CSRF, XSS). 
These issues, which stem from an HTTP-based architecture, can be exploited more readily in scenarios where notebooks are exposed to broader networks or multi-user environments. 
In practice, many organizations deploy notebook servers on corporate networks, in the cloud, or as part of shared platforms—environments that expand the attack surface beyond an individual's machine.

One key tension is the need to balance usability and security. A subset of security-related commits intentionally relaxes security controls to address the complexity of managing TLS certificates or cross-origin settings, particularly when notebooks are used on non-standard platforms or by non-professional users. 
While these relaxed settings simplify local installation and reduce support overhead, they pose considerable risks if the same configurations are adopted in production. 
For instance, disabling strict authentication or cross-origin checks could enable unauthorized access or session hijacking if the server is exposed to a public network.

These findings emphasize that notebook infrastructures are not intended for large-scale or security-critical contexts. 
Security-conscious deployments should implement robust authentication, enforce TLS for all traffic, apply strict cross-origin policies, and keep dependencies updated. 
Additional measures such as containerization or sandboxing can further restrict the scope of potential exploits. 
Nevertheless, since many Jupyter installations remain local, developers are facing a dilemma: adding robust security controls often introduces additional complexity, which can harm the out-of-the-box experience that makes Jupyter Notebook usable.

\smallskip\finding{\textbf{RQ3 Summary:} Jupyter Notebook infrastructures are prone to common web-based vulnerabilities. We also observed a tension between usability and security, where certain commits relaxed default protection settings to improve ease of deployment, especially in single-user scenarios. }

\section{Threats to Validity}
\label{sec:threats}

\smallskip\noindent\textbf{Conclusion validity:} Bugs can be subjective without a clear specification of how a piece of program should behave. 
Therefore, there is a potential threat regarding the identification and classification of the bugs in our study.
To mitigate this issue, the classification was done in a systematic process in multiple iterations and verified by several authors.

\smallskip\noindent\textbf{External validity:}
Despite the large size of our datasets, the notebooks used in our evaluation come from a relatively homogeneous source (e.g., GitHub and Kaggle).
Thus, the results on notebooks from other sources (e.g., company internal repositories or specialized communities) might be different.
However, the data set should be reasonably representative for most real-world Jupyter notebooks given that GitHub and Kaggle are two major platforms for hosting notebook-related resources and they contain a wide range of topics.

\smallskip\noindent\textbf{Construct validity:}
If one category of certain bugs escaped our iterative sampling, then our approach would fail to include this bug in our bug taxonomy. 
We mitigate this issue by resampling the data for our manual analysis until saturation.
\section{Related Work}
\label{sec:related}

\smallskip\noindent\textbf{Empirical Studies on Jupyter Notebooks:}
Much work has been done investigating software engineering practices for data science in Jupyter Notebooks. 

The closest work is the concurrent study done by De Santana et al.~\cite{desantana2024jupyter} that analyzes Notebooks from GitHub and Stack Overflow posts, and conducts interviews with Jupyter developers in order to develop a taxonomy of problems related to Jupyter Notebooks. 
While their study is thorough, it focuses on higher-level issues encountered by developers (e.g., the kernel crash, the notebook cannot be converted to another format) while we focus on bugs in the source code (e.g., undeclared variable, error in test code, incorrect API code). 
Our lower-level taxonomy complements previous work well, and the connection to actual source code makes it more actionable than the previous taxonomy.
In addition, while they filter tutorials and projects related to courses or books, this was done automatically, leading to potentially mislabelled repositories while we manually went through over 1,000 notebook repositories to ensure the quality of our dataset. 
Finally, our inclusion of Kaggle Notebooks, an independent secondary source of Notebooks further complements De Santana et al.~\cite{desantana2024jupyter} contributions.

Previous work~\cite{dong2021splitting} presents a qualitative study of cleaning activities in Jupyter Notebooks (adding, removing cells, etc.).
~\cite{kallen2021jupyter,ritta2022reusing} analyses \rawnumber{2.7} million Jupyter notebooks hosted on GitHub and found that \rawnumber{70\%} of code snippets were clones and \rawnumber{50\%} of Notebooks have no unique code snippets. 

Grotov et al.\cite{grotov2022large} quantitatively compare Python scripts and Jupyter Notebooks, finding that Notebooks have more stylistic issues—suggesting different developer behaviors and the need for Notebook-specific studies. In contrast, Adams et al.\cite{adams2023comparison} report that Notebooks used in machine learning have fewer stylistic issues than Python scripts. Nalin~\cite{patra2022nalin} explores variable name/value inconsistencies in Notebooks using AI and dynamic analysis. Pimentel et al.\cite{pimentel2021understanding,pimentel2019large} and Wang et al.\cite{wang2020assessing,wang2020better} focus on Notebook reproducibility, proposing tools and techniques to detect and resolve related issues. 
Unlike these studies, our work provides an analysis of bugs and security vulnerabilities in the Jupyter Notebook ecosystem, combining both quantitative and qualitative perspectives.

Quaranta et al.~\cite{quaranta2022eliciting} interviewed Notebook developers and studied best practices in 1,380 Jupyter Notebooks from Kaggle. They found that experts are generally aware of best practices (using version control, testing, etc.) but inconsistently apply them. 
Settewong et al.~\cite{settewong2022visualize} investigated how visualization is used in competition notebooks to explain coding solutions and proposed a taxonomy of 9 types of visualizations used by expert notebook users. Chattopadhyay et al.~\cite{chattopadhyay2020s} identify 9 pain points of computational notebooks. These pain points are not directly related to bugs in the code and do not overlap with our taxonomy. Finally, 
Van Binsbergen et al.~\cite{van2020principled} survey Read-eval-print-loops principles, which are used by computational notebooks.

\smallskip\noindent\textbf{Jupyter Notebooks tools and datasets:}
Quaranta et al. proposed a large dataset of Jupyter Notebook, KGTorrent~\cite{quaranta2021kgtorrent} to help researchers. 
Pynblint~\cite{quaranta2022pynblint}, NBLyzer~\cite{subotic2022static}, and Julynter~\cite{pimentel2021understanding} are existing static analysis tools for Jupyter Notebooks. While being a step in the right direction, they are still lacking most standard linting features. Merino et al.~\cite{merino2022making} discuss the possibility for software engineers to develop widgets to increase users' access to the internal states of the executed notebook.  Other extensions proposed in previous work~\cite{cunha2021context} help determine which cells should be migrated, reducing the notebook's states and increasing performance.

\smallskip\noindent\textbf{Bug Taxonomy:}
Many bug taxonomies have been proposed, focusing on different ecosystems such as autonomous vehicle bugs~\cite{garcia_comprehensive_2020,wang2021exploratory}, deep learning systems~\cite{humbatova_taxonomy_2020,shen2021comprehensive,jia2020empirical,sun2021experience}, JavaScript~\cite{gyimesi2021bugsjs,hanam2016discovering}, HTML~\cite{macklon_taxonomy_nodate}, video games~\cite{lewis_what_2010}, Python {API}~\cite{hu_empirical_2023,kamienski2021pysstubs}, Data Analytics~\cite{ahmed_characterizing_2023}, test code~\cite{vahabzadeh2015empirical}, blockchain~\cite{wan2017bug}, internet of things~\cite{makhshari2021iot}, infrastructure as code~\cite{rahman2020gang}, open-source software~\cite{tan2014bug,catolino2019not,valdivia2018characterizing,zhao2017towards}, compiler~\cite{romano2021empirical}, regular expressions~\cite{wang2022demystifying} and security bugs~\cite{wei2021comprehensive,jimenez2016empirical}. While they follow a similar process as ours, all these taxonomies are in widely different domains, making these works very different from the current paper.
\section{Implications \& Conclusion}

\noindent\textbf{Practical Implications.}
Our study offers several practical insights for software developers, data scientists, and AI engineers involved with Jupyter Notebooks. First, given that configuration-related issues were found to be among the most common bugs, practitioners should adopt rigorous configuration management strategies, including clearly specifying library versions and paths, maintaining environment files, and leveraging containerization solutions such as Docker. Integrating notebook-specific linting and static analysis tools can further reduce frequent mistakes, such as API misuse or typos.

From a research perspective, our taxonomy of notebook bugs and the insights derived from analyzing vulnerabilities highlight the need for further studies focused on tailored software engineering methodologies specifically suited for computational notebooks. Future research could explore effective notebook-specific debugging techniques, develop more sophisticated static analysis tools that understand the notebook's cell-based execution model, or investigate notebook-friendly refactoring methods. Lastly, given the rising prevalence of security vulnerabilities within notebook deployment frameworks, researchers should prioritize developing notebook infrastructure that better balances usability with security, perhaps through context-aware adaptive security mechanisms or enhanced user awareness and education initiatives.

\smallskip\noindent\textbf{Conclusion.}
We proposed a large-scale empirical study, including a quantitative analysis of the notebook ecosystem, a qualitative analysis of bugs in notebook documents, and a study of vulnerabilities in notebook frameworks. We studied \totalGitHubSE Jupyter notebooks from GitHub and \totalKaggleNotebooksParsed notebooks from Kaggle. Based on our analysis, we deduce that configuration issues and API misuse were two of the most common errors that notebook users faced and presented a new taxonomy for bugs faced by users working in Jupyter notebooks.

Overall, our work highlights that attractive features of Notebooks such as interactivity come at a cost, increasing configuration issues and raising concerns about the reproducibility, maintainability, and security of notebook projects.

\smallskip \noindent \textbf{Data Availability:} Our replication package and dataset are available on our anonymous GitHub\footnote{\url{https://github.com/jwyjohn/Exploring-the-Jupyter-Ecosystem}}.

\bibliographystyle{spmpsci}
\bibliography{manuscript} 
\end{document}